\documentclass{article}

\usepackage[accepted]{vietnam}

\usepackage{natbib}
\usepackage{wrapfig}
\usepackage{graphicx}
\usepackage{cite}

\def\be{\begin{equation}}
\def\ee{\end{equation}}
\def\bea{\begin{eqnarray}}
\def\eea{\end{eqnarray}}

\def\Msun {\hbox{$M_\odot$}}

\def\HII  {\hbox{H{\sc ii}}}

\def\UCHII  {\hbox{UCH{\sc ii}}}

\def\apj{{ApJ} }
\def\apjl{{ApJL} }

\def\aap{{A\&A} }
\def\mnras{{MNRAS} }
\def\araa{{ARA\&A} }

\def\apjs {{ApJS}}
\def\apss {{Astrophys. and Space Sc.}}

\def\nat {{Nature}}

\begin{document}
\vspace*{4cm}

\title{(Mostly) Observational Aspects of High-Mass Star Formation}

\author{Peter Schilke}{schilke@ph1.uni-koeln.de}
\address{I. Physikalisches Institut der Universit\"at zu K\"oln, Z\"ulpicher Str. 77, 50937 K\"oln, Germany}


\begin{abstract}
A review on current observations of high-mass star formation is given, with a little bit of theoretical background.  Particular emphasis is given to the, in my opinion, most important observations to put strong constraints on models of high-mass star formation: the existence and properties of high-mass starless cores, the existence or not of isolated high-mass stars, the possible support mechanisms of starless cores, the role of filaments in the mass transport to high-mass cores, ways of characterizing cores, the binary properties, and the properties of disks around high-mass stars.   
\end{abstract}

\section{Introduction}

\begin{figure} 
\centering
\includegraphics[width=0.5\textwidth]{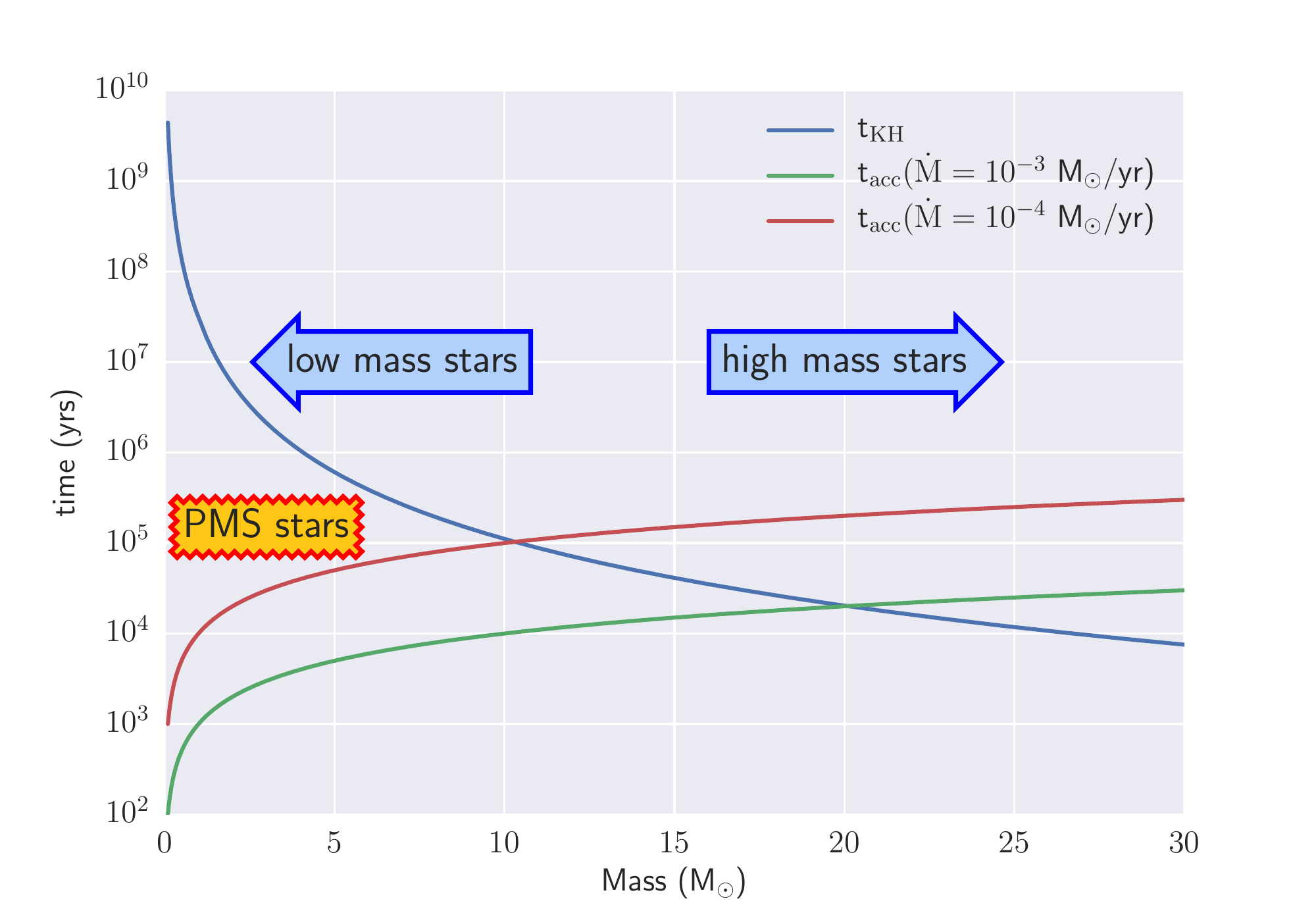}

\caption{Kelvin-Helmholtz time scale vs. accretion time scale of high mass stars vs. low mass stars, for varying accretion rates.  Only low mass stars have a pre-main sequence phase, where they have stopped accreting, but get their energy from gravitational contraction. 
} \label{schilke:fig1}
\end{figure}

High-mass stars dominate the energy input in galaxies, through mechanical (outflows, winds, supernova blast waves) and radiation (UV radiation creating  \HII\ regions) input.  Particularly through supernova explosions, they can shape whole galaxies \citep{Bolatto2013}. They also enrich the ISM with heavy elements, which in turn modifies the star formation process. Yet, their formation process will differ from low-mass stars in significant ways: while the Kelvin-Helmholtz timescale of low-mass stars is significantly longer than the time required to assemble them, for any reasonable accretion rate it is shorter for high-mass stars (Fig.~\ref{schilke:fig1}).  This has as a consequence that high-mass stars above a certain mass (again depending on the accretion rate) will continue accreting after reaching the main sequence.  These still deeply embedded stars are mostly (except in some favorable geometries) invisible at optical and near IR wavelengths, which limits the observational means of characterizing them to mid-IR to cm wavelengths.  Also, their radiative feedback in the later stages of accretion is considerable.  \citet{WolfireCassinelli1987} show that spherical accretion with normal grains could make the creation of stars of more than 10 \Msun\ impossible, because the radiation pressure on dust would halt accretion.  As stars more massive than 10 \Msun\ do exist, this led to various attempts to circumvent this barrier.

\section{Theories of High-Mass Star Formation}

\citet{WolfireCassinelli1987} speculate about different dust properties. A later insight was to abandon the assumption of spherical accretion.  Any angular momentum of the infalling gas will lead to the creation of a disk, which funnels accretion in the equatorial plane with much higher rates per area.  At the same time, most of the radiation escapes in the polar regions, and therefore cannot interact with the infalling dust particles.  This was first investigated by \citet{YorkeSonnhalter2002}.

\subsection{Turbulent core vs. competitive accretion}



But even with spherical accretion, the radiation pressure barrier can be overcome if the accretion rates are high enough (see Fig.~5 from \citealt{WolfireCassinelli1987}). If the accretion rate depends only on thermal support, \citet{Shu77} find typical values of 10$^{-5}$ \Msun/yr, which is not sufficient for forming high-mass stars.  \citet{McKeeTan2003} therefore considered turbulent support, which allows much higher accretion rates.  They call their model ``turbulent core'', but it is also known as ``monolithic collapse''.   Part of its appeal is that it is analytical, so it can make many easily accessible predictions on derived values (e.g.\ mass vs. time, or density profiles as function of time).  It does however make some quite strong assumption on initial conditions. Apart from the spherical approximation, it starts with a strongly peaked density distribution ($n\propto r^{-1.5}$). It is therefore fully applicable if and only if such strongly peaked cores exist. One characteristic of this model is that the final stellar mass is pre-assembled in the collapsing core (even if it may form binary or multiple stars), i.e.\ it assumes that the clump mass is isolated from the rest of the cloud and directly maps into the initial stellar mass distribution.  

An alternative approach was taken by \citet{Bonnell01}.  They considered star formation in a cluster.  All stellar embryos are created with equal mass, and then gather mass through Bondi-Hoyle accretion.  Their accretion rate therefore is determined by their location in the cluster potential -- near a gravitational well, where more mass is available, a star can accrete more, in the outskirts of the cloud, less.   Gravitational interactions between stars can kick them out of rich feeding zones, which terminates their accretion.  Since the stars have to compete for their resources, this model is called ``competitive accretion''. It uses a uniform gas density as initial condition, and the original model was isothermal.  In this model, the final star can, depending on its trajectory, draw from the vast mass reservoir of the whole cloud.  Here, there is no connection between the mass of its birth core and the final stellar mass.

Turbulent core and competitive accretion were almost orthogonal approaches to the problem of high-mass star formation, and to a detached observer, they shared the same type of pitfalls: both used very specific and somewhat arbitrary, but very different, initial conditions (although they were physically motivated), and both lacked important physics, mostly related to feedback.  Both of them have been updated with new physics in the last years, and my impression is that their only real difference these days is the choice of the initial conditions.  In the following, I will refrain from discussing the new physics in detail (magnetic fields, mechanical and radiative feedback, both thermal and ionizing), since these will be covered by other talks.  However, I will briefly mention the question of initial conditions, since those can be tested observationally.  Some more recent papers that go toward merging the paradigms are \citet{LiNakamura2006, Fall2010, NakamuraLi2011, MurrayChang2012, Zamora2014, MatznerJumper2015, LeeHennebelle2016}, but this list is not exhaustive.

\subsection{Initial Conditions}
The initial conditions of the two model classes are very different: while the turbulent core model used a highly peaked density structure, the density structure of the molecular cloud of the competitive accretion model was flat. It seemed plausible that these different intial conditions had influence on the results, and this was systematically investigated by \citet{Girichidis2012}. They ran the same models with the same physics, but different initial conditions, and investigated the resulting cluster structure.  Indeed it was found that peaked distributions resulted in no sub-clustering (just like turbulent core models), while flat density distributions resulted in strong sub-clustering (just like competitive accretion).  

Therefore, one of the strongest preconditions for getting realistic models of high-mass star formation is to get the initial conditions right.  This is not trivial, because the initial conditions of star formation are the results of cloud formation models, which in turn depend on their own initial conditions, determined by galaxy formation and evolution models, etc. Since these models all deal with vastly different scales (from kpc in galaxy models to AU for star formation), it is numerically not possible to calculate this all self-consistently. Instead, one would calculate models on one hierarchical scale, and use this as input for the next smaller scale.  There are a couple of simulations in the literature that lend themselves to such zoom-in studies: the SILCC simulation that calculates a large section of a galactic disk including supernova feedback, but without a galactic potential (i.e. no spiral arms; \citealt{Walch2015}); other simulations do use a spiral potential, but limited physics, particularly no feedback \citep{Dobbs2013, Dobbs2015, Smith2014}. 

\section{Observations of High-Mass Star Formation}
The ultimate way to determine the initial conditions is to use observations.  This poses technical problems though: apart from the ubiquitous one that one observes just a 2-d projection in space and a 1-d projection in velocity, while one needs to reconstruct a 3+3d phase space, high-mass star formation is rare and short-lived.  Thus, there are only a very limited number of instances at any given time in a galaxy, and they are, on average, far away.  Moreover, since high-mass star formation happens deeply embedded, only observations at FIR and longer wavelengths can penetrate the cores.  Observations hence require high-resolution instruments in the mm/submm wavelength range.  Because of the long wavelengths compared to the optical, these instruments must be huge to achieve the necessary spatial resolution.  Only recently has, with ALMA, such an instrument become available.  ALMA has already started to change our view of star formation,  and we can certainly expect many new results in the future.

\subsection{High-Mass Starless Cores}

One of the testable predictions is based on the fact that the precondition for the turbulent core model is the existence of high-mass prestellar cores, i.e.\  objects that have pre-assembled masses sufficient to form high-mass stars in ($> 40 M_\odot$ for the lower end of high-mass stars), but have not formed stars yet.  Are there such objects, and are they common?  

\citet{Motte2007} found 129 dense cores in Cygnus X, among them 
40 with masses $> 40 M_\odot$.  Of those, 17 were found to be IR quiet, i.e.\ show no sign of a star having been formed already in the IR.  \citet{Bontemps2010} observed the 5 brightest of these objects and found that all but one are sub-fragmented, that is, they are not forming high-mass stars, but a small cluster of lower mass stars.  The one remaining candidate was shown by \citet{DuarteCabral2013} to have an outflow.   Therefore, star formation is already ongoing, and this core has to be removed from the list of high-mass prestellar mass candidates. In an ongoing more complete SMA survey of the Cygnus X region, preliminary results seem to support this trend: most cores fragment, and most cores already have outflows (Keping Qiu, priv. comm.)  

In another region, \citet{Tan2014} found a candidate high-mass core of 60\ \Msun, dark at 70 $\mu$m, which would need to be supported by strong magnetic field to be stable.  \citet{Kong2015} corroborates a slow collapse of this cloud, using deuterated molecules as a chemical clock.  Thus, in 2015 this seemed to be a very good candidate for a starless high-mass core.  In 2016 however, both \citet{Tan2016} and \citet{Feng2016} found an outflow in this core, which indicates that star formation has already begun, and thus the source has to be removed from the list of candidates for high-mass prestellar cores. The best candidate to date that I am aware of has been found by \citet{Cyganowski2014}, who reports on a very dense core of $\approx 30 \Msun$ showing no sign of star formation.  Based on these observations, which do not represent statistically relevant samples, at present no convincing evidence of genuine high-mass prestellar cores, as required by the turbulent core model, has been found.  Because of the scarcity of the samples, the existence of such objects cannot be excluded though.  They do seem to be exceedingly rare, the known candidates not extremely high-mass (assuming a star-formation efficiency of 30\%, the \citet{Cyganowski2014} source would form at most a 9 \Msun\ star, and more likely a multiple system with lower mass stars, or even a small cluster.   

This is a very limited sample though, and by now there are large surveys of the galactic plane, which sample many more sources.  And indeed, \citet{Csengeri2014} find, in their ATLASGAL sample, that 25\% of their high-mass cores are IR dark.  However, this is based on data taken with a rather large beam, and no test for fragmentation or outflows as early sign of star formation has been performed yet, so this is a strict upper limit.  Systematic studies of these candidates with ALMA at high resolution to establish the fragmentation state and to search for outflows as early signs of star formation are thus urgently needed to put the findings on a solid statistical basis.

One could approach the problem from the other end.  While most high-mass stars do form in clusters, high-mass stars that form in isolation, i.e.\ without an accompanying cluster of smaller stars, would point to formation out of quiescent, density peaked condensations, as these are required for monolithic collapse.  There are many isolated O-stars, but the majority of them have been ejected from a cluster by stellar 3-body interactions, so one carefully has to exclude them being runaways.  Recent searches have been conducted by \citet{Bressert2012} in the 30 Dor region in the LMC and \citet{Oey2013} in the SMC.  They find 15 and 14 isolated OB stars, respectively, that are not runaways to the best of their ability to establish that.  Earlier, \citet{deWit2005} finds that, in our Galaxy, $4\pm2$\% of high-mass stars are found in isolation, without being obvious runaways.  One has to be careful though, since slow runaways are possible \citep{Banerjee2012}, which means that, if they are ejected early in the formation process, but after they have acquired their mass, those objects can reach considerable distances from their origin without betraying their runaway status by high systemic velocities.  And, finally, \citet{Stephens2017} find that  many allegedly isolated high-mass stars are, on closer inspection, surrounded by clusters of lower mass stars after all.  One basically reaches the same conclusion here:  high-mass star formation in isolation seems to be rare or nonexistent, but the current data do not allow to rigorously exclude the existence of such objects.

However, the real question is not if monolithic collapse exists, but if it is a common mode of high-mass star formation.  I feel this question can already be answered with the current knowledge: monolithic collapse out of pre-assembled centrally peaked high-mass prestellar cores does not constitute the dominant mode of high-mass star formation.  The reason for this is that the mechanisms to assemble cores seem to be unfavorable to producing a core with the necessary initial conditions for monolithic collapse, without igniting star formation along the way.

\subsection{Support mechanisms}
Since thermal pressure at the temperatures of most molecular cores is insufficient for support against gravitational collapse, turbulence  \citep{McKeeTan2003}, or magnetic fields \citep{Li2014} have been evoked as support mechanisms.  The timescales of star formation are still not very clear, but evidence suggests that it does not proceed at free-fall \citep[see discussion in][]{Dobbs2014}, so some support must exist.  A thorough discussion of all this is beyond the scope of this review, but I would like to point out two things:  turbulence is not producing an isotropic pressure, so while it can provide support to parts of a cloud, there will be channels where inward movement of gas is not inhibited.  Since therefore there are large deviations from spherical symmetry \citep[see e.g.][]{Smith2013}, any measured infall rates etc.\ will have to be taken with caution, and should be interpreted only in a statistical sense.  

Magnetic fields generally seem not to be strong enough to fully support cores \citep{Crutcher2012}, although they do modify the dynamics, because even for fields too weak to halt collapse movement along the field lines is still easier than perpendicular to it.  The observation of a polarization hole toward cores, which could be caused by tangling of the magnetic field on small scales \citep{Hull2014}, and the observed random alignment of outflows and magnetic fields \citep{Offner2016} hint at magnetic fields not dominating the dynamics at small scales.  These results, however, were obtained for lower mass cores.  The few observations of polarization toward high-mass cores \citep{Tang2013} support the picture that gravity dominates the magnetic field at small scales.  More observations, at high resolution, of both magnetic field directions through dust polarization and magnetic field strength from Zeeman-splitting are necessary however, to assess the range of importance magnetic fields have in different high-mass star-forming regions.

\subsection{Finding High-Mass Protostars}

Early searches for high-mass protostars used IRAS color-color criteria and then targeted searches for cm continuum radiation from embedded \UCHII\ regions \citep{WoodChurchwell1989}. \UCHII\ regions are readily observable signposts for high-mass stars that already have reached the main sequence, and probably already have accreted most of their mass.  Many earlier stages were found adjacent to the \UCHII\ regions, mostly in the form of hot cores \citep[for a review see][]{Cesaroni2005Rev}. This however was a very biased search, and one can ask the question how many high-mass protostars were missed, since they did not emit at cm wavelengths.  \citet{Sridharan2002} took the approach to look for sources with the same IRAS colors as \citet{WoodChurchwell1989}, but explicitly requiring the absence of cm emission above a certain level.  The resulting sample of High Mass Protostellar Objects (HMPOs) indeed proved to be a somewhat earlier evolutionary stage, although deeper cm searches showed cm emission in many of them, on a lower level. 
\begin{figure}
\centering 
\includegraphics[width=0.5\textwidth]{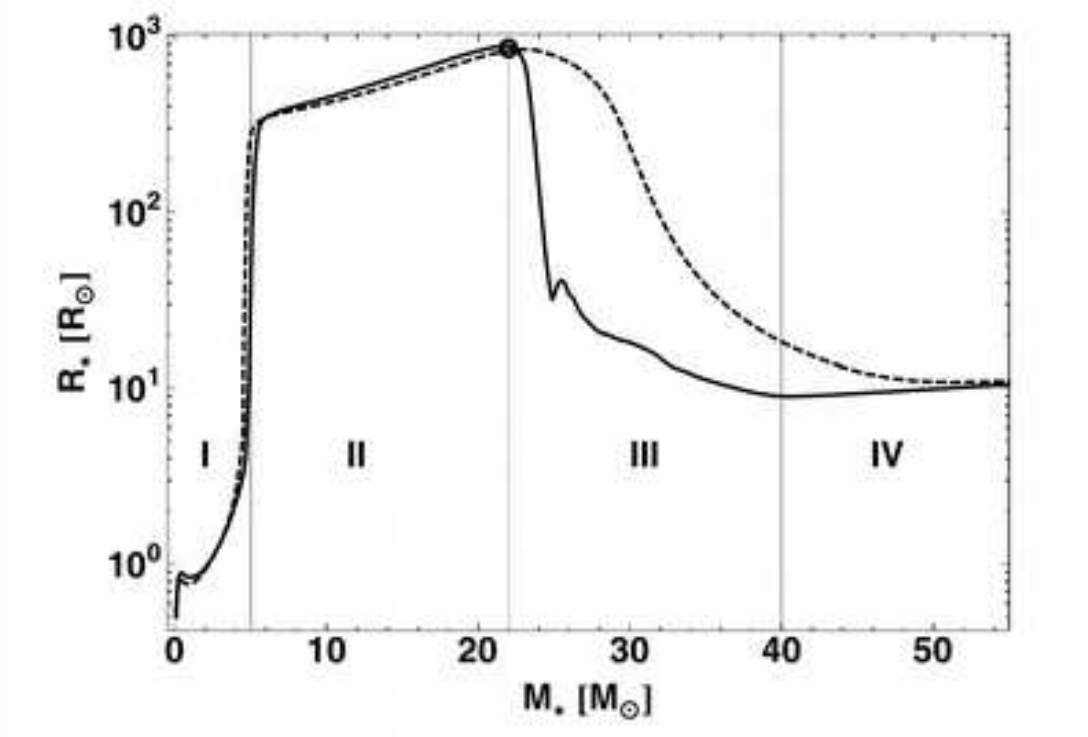}
\includegraphics[width=0.5\textwidth]{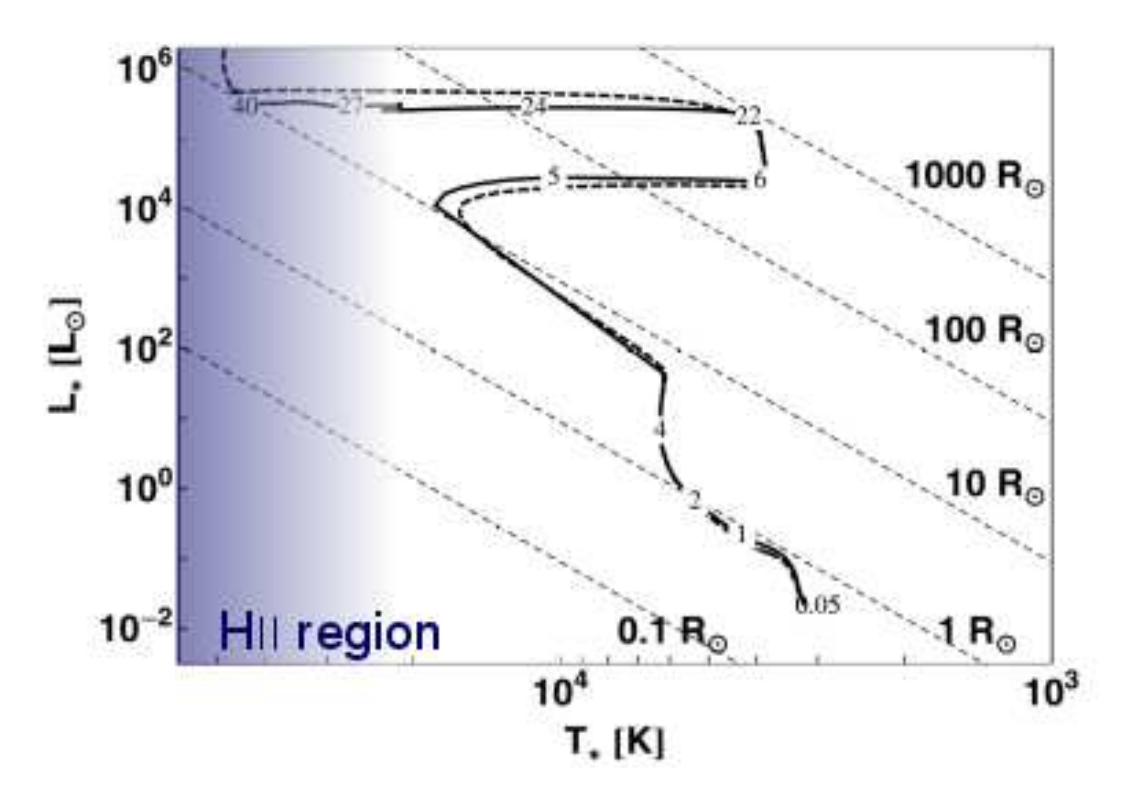}

\caption{Radii of stellar evolution (top) and evolutionary tracks  (bottom) (Fig.~4 and 1, respectively, from \citealt{Kuiper2013}, based on calculations by \citealt{HosokawaOmukai2009} and \citealt{Hosokawa2010} \copyright AAS. Reproduced with permission).} \label{schilke:fig3}
\end{figure}

One can ask what theory has to say about the properties of high-mass protostars, and in particular at which stage one can expect to detect \UCHII. If one assumes that high-mass protostars look just like ZAMS stars, one would expect \UCHII\ to appear when the stars is of spectral type B2, or about 10~\Msun\ \citep{Straizys1981}.   However, the protostars are still accreting mass, and hence do not look like ZAMS stars.  \citet{HosokawaOmukai2009} and \citet{Hosokawa2010} have calculated the stellar structure for high accretion rates ($\approx 10^{-3}$\Msun/yr), and find that for a long time during their development, the protostellar radius is very large, and the protostars are bloated (Fig.~\ref{schilke:fig3}). This means that, although they have a high luminosity, their temperature is too low to emit a sufficient amount of EUV photons, and the \HII\ region does not appear until they have reached about 30~\Msun.  Thus, the observational absence of free-free radiation is not necessarily evidence for the absence of a high-mass protostar.

Since column densities, and thus extinctions, can be extremely high, to the point of blocking 70 $\mu$m emission for a protostar \citep{Tan2016, Feng2016}, outflows in CO or SiO may be the best observational signpost of star formation.  Both tracers pose problems though:  while it is uncontested that SiO in most cases is produced by shocks \citep{Schilke1997SiO, Gusdorf2008} (but see \citealt{Schilke2001} for SiO from PDRs), and is usually uncontaminated by ambient emission, it is by no means obvious that this shock is produced by an outflow.  Occasionally cloud-cloud collisions have been evoked as source of SiO \citep[e.g.][]{Louvet2016}.  There also are outflows without SiO \citep{Beuther2002}, although in general SiO seems to have a high detection rate in high-mass clumps \citep{Csengeri2016}.  Narrow SiO may trace fossil shocks \citep{JimenezSerra2010}. Low-J CO on the other hand suffers from contamination by ambient gas, particularly for outflows in the plane of the sky, and there is also contamination from PDR heated gas and turbulent dissipation \citep{Pon2012} even in mid-J lines. Unambiguous assignment of an outflow as origin of wide wings, or SiO emission thus is not possible without verifying the morphology.  Mapping instruments such as ALMA again will excel in this task.

Comparison of observations with star formation models, which of course do not model any specific source, requires statistically relevant source samples, but also good modeling of the observations to be able to extract obervable properties that allow to distinguish between prediction from different model types.  Due to large surveys e.g.\ by Herschel, APEX or IRAM and in the future by ALMA, the data will be available.  What often is lacking is a comprehensive and fast modeling of the often multi-wavelength data, to extract the maximum of information from the data.  There are steps in the right directions, \citep[see e.g.][]{Schmiedeke2016}, but still work needs to be done to accelerate and automatize the data analysis. We also need to develop more and better metrics for statistical comparison of observations and data. It is to be expected that machine learning methods will play an increasingly important role in the future.

\subsection{Filaments and Mass Flow}
One of the lasting legacies of \emph{Herschel} is the result that filaments are ubiquitous in the interstellar medium, and that star-forming cores are often found at the intersection of filaments\citep{Schneider2012}. Thus, it was suggested and then shown \citep{Peretto2013, Liu2015} that mass flows along the filaments toward the cores.  Observing this is not trivial, since nothing is known about the orientation of the filaments on the plane of the sky, and velocity gradients cannot be unambiguously connected to infalling or out-streaming gas.  The traditional method of using line profiles to detect mass infall \citep{Evans1999}, which was derived for spherical infall, has been shown by \citet{Smith2013} not to provide unambiguous results, depending on the geometry. 

\subsection{Binary Fraction}
One observable in high-mass stars is the binarity or general multiplicity fraction, which is close to 100\% for high-mass stars \citep{DucheneKraus2013}.  Turbulence will create multiple seeds in a core, which can then form individual stars \citep{Offner2010}.  Turbulent fragmentation will be suppressed by magnetic fields  \citep{Hennebelle2011} or thermal feedback \citep{Krumholz2007, Bate2009}. Multiple systems formed by this method will have rather wide separations, although dynamical interaction or accretion can diminish the separation (harden a binary, \citealt{Bate2002}).  This process tends to form equal mass binaries.  Disk fragmentation \citep{Kratter2006} will lead to tight binaries, but with unequal mass ratios, since the star condensed out of the disk will have lower mass than the original central star.  Since most stars, and particularly high-mass stars, form in clusters \citep{Bressert2010}, one has to consider not only interaction after star formation, but also before.  There is evidence that the protostellar disk sizes can be truncated by encounters in dense clusters \citep{Vincke2016}, which also would influence the mass flow onto the star, and the final stellar masses.  Thus, a variety of mechanisms can influence the binary fraction as well as the distribution of the mass ratios and orbital parameters, but this will change with time through interactions, and if constraints on star formation are to be derived, one has to observe all these parameters in a very early stage of development.  

\subsection{High-Mass Disks} 
As already mentioned, high-mass star formation theories predict accretion disks to sustain accretion in the presence of radiation pressure.  Determining the disk properties observationally thus gives important constraints on models.  The predicted sizes are between 100 and 1000 AU \citep{Kuiper2015}.  There have been reports on disks around B-stars \citep{Cesaroni2005, Cesaroni2014, SanchezMonge2013}, which report sizes of about 2000~AU, and Keplerian rotation.  When the resolution is high enough to resolve the disk, asymmetries are found, attributed to tidal interactions with a central binary \citep{Cesaroni2014}.  

Around O-stars, prior to 2014, only very large rotating structures called toroids were found \citep{Beltran05, Beltran2011}.  These structures are large (10,000 AU), massive (100\Msun) and rapidly contracting, i.e.\ not showing Keplerian rotation.  They are supposed to feed clusters rather than single stars. \citet{Hunter2014, Zapata2015, Johnston2015} and \citet{Ilee2016} present disk candidates around O-stars, with \citet{Johnston2015} and \citet{Ilee2016} showing the best evidence for Keplerian rotation.  The inferred sizes (1000-2000 AU) are at the outer range of expected sizes, but do at present not exclude the possibility that these actually are circumbinary disks.  This would not be surprising, since \citet{Sana2014} find that about 80\% of O-stars have a companion closer than 1000 AU. It is to be expected that, with the new ultra-high resolution capabilities of ALMA, light will be shed on this question.  

It would appear that the easiest tracers for the existence of disks are outflows, since they generally are easy to detect, and indeed are found to be ubiquitous in high-mass star-forming regions. One might hope that studying such flows could allow to derive some properties of the disk of origin, i.e.\ the infall rates, the ejection mechanism, the energies etc. \citet{Beuther2005} propose an evolutionary sequence of outflows from massive stars, with older stages having larger opening angles.  \citet{Kuiper2016} explain this by the onset of strong radiation pressure in the outflow cone. However, high-mass stars form in clusters, which also contain low-mass stars with outflows. With time, outflows grow to large sizes, which means that in this kind of environments they tend to overlap, at least in projection. Hence it is very difficult if not impossible to observationally isolate a specific outflow, derive its properties, or even unambiguously assign its origin to a specific source.  Examples in a relatively sparse clusters are shown in \citet{Beuther2002, Beuther2003}, and a theoretical investigation of how aligned multiple flows can mimick a higher-mass flow is discussed in \citet{Peters2014}.  The conclusion is that even in the age of ALMA, where at least the instrumental resolution is sufficient to separate flows, which was not the case in the past, the instrinsic properties of multiple flows make deriving disk properties from outflow properties a hazardous enterprise, which has to be done with extreme care.

\section{Outlook}
The properties most important to study observationally, and that will give the strongest constraints on models (in my biased view) are the following:

\begin{itemize}
\setlength\itemsep{0.0pt}
\item \textbf{Are high-mass cores connected to a larger cloud mass reservoir?}  This can be studied through observations of mass-flow along and onto filaments on both large and small scales.
 \item \textbf{What do the initial conditions look like -- what is the fragmentation status, what are the density profiles?}  ALMA mapping of high-mass cores and sophisticated modeling of the results will be necessary.
 \item \textbf{What is the role of magnetic fields in shaping clouds and influencing dynamics?}  This will require dust polarization studies at large and small scales, and Zeeman splitting observations.
 \item \textbf{Are there disks around high-mass stars and what do they look like?}  Here, high-resolution ALMA observations will advance the field within a very short time.
 \item \textbf{What are the binary/multiple properties of high-mass stars at early stages?}  High-resolution observations with ALMA, of the embedded stages, and IR/IR Interferometric observations at somewhat later stages, will shed light on this.  Since this will involve observations of dense clusters where chance projections can be mistaken for multiplicity, a careful statistical analysis is needed.
\end{itemize}
A  word of caution:  given the complexity of the star formation process, one does not necessarily expect to that it proceeds in the same way everywhere.  It is well probable that there are some cores that do resemble the initial conditions for turbulent cores, others may more look like the ones in the competitive accretion scenario.  In some places, magnetic fields may be important, in others, not.  So the ultimate goal would be to first describe the properties discussed above on a statistically sound basis, and then trying to find the causes for the properties being what they are.  There seem to be some invariants (e.g.\ the IMF) that do not seem to be very sensitive to the details of star formation.  Explaining these quantities form necessary conditions for star formation theories:  mechanisms that do not produce the observed IMF can be discarded.  They are not sufficient criteria though to discriminate among the family of theories that do manage to reproduce them.  Thus, to connect theory and observations, 
\begin{itemize}
\setlength\itemsep{0.0pt}
\item theories have to identify strong predictions of specific models that can be tested (falsified) by observations.  Examples would include stellar rotation rates, magnetic fields, multiplicities, companion rates, cluster kinematics, etc.
\item observers and theorists have to develop metrics beyond the standard ones (like probabulity distribution functions, power spectra etc.) to compare theory/simulations and observations, e.g.\
to characterize filament properties, distribution of widths, lengths, velocity gradients, topology (branching properties) etc.
\item as a necessary first step, theories have to provide mock observations (as the ISM people do already), since in particular molecular line observations
\begin{itemize}
\item show a chemically filtered parameter space (i.e.\ specific molecules exist only in specific environments).  This has advantages and disadvantages, depending on the availability of a tracer for the tested property,
\item reduce the dynamic range of traced conditions because of radiative transfer effects (high opacities), and detection thresholds.
\end{itemize}
\end{itemize}

\section*{Acknowledgments}
This work has been supported by the Collaborative Research Centre 956, sub-projects A6 and C3, funded by the Deutsche Forschungsgemeinschaft (DFG) and by the German Ministry of Science (BMBF) through contract 05A11PK3. I thank Chris Matzner for a critical reading of the manuscript and helpful comments.

\setlength{\bibsep}{0.0pt}

\end{document}